\documentclass[conf]{new-aiaa}
\usepackage[utf8]{inputenc}
\usepackage{comment}
\usepackage{graphicx}
\usepackage{amsmath}

\usepackage[version=4]{mhchem}
\usepackage{subcaption}
\usepackage{float}
\usepackage{pgfplots}
\usepackage{tikz}
\usetikzlibrary{positioning}

\usepackage{booktabs}
\usepackage{dirtytalk}

\usepackage{siunitx}
\usepackage{longtable,tabularx}
\usepackage[dvipsnames]{xcolor}

\title{Efficient Sensor Fusion Through Covariance-Constrained Observation Decimation (CCOD) }
\author{Andres Enriquez Fernandez\footnote{Graduate Research Assistant, Department of Aerospace and Mechanical Engineering, AIAA Member.}}
\affil{The University of Texas at El Paso, El Paso, TX}
\author{Erik Blasch\footnote{Program Officer, AFRL, and AIAA Fellow}}\affil{AFRL, Rome, NY}
\author{Angel Flores-Abad\footnote{Associate Professor, Department of Aerospace and Mechanical Engineering, AIAA Member.}}
\author{John J. Bird\footnote{Assistant Professor, Department of Aerospace and Mechanical Engineering, AIAA Member.}}
\affil{The University of Texas at El Paso, El Paso, TX}

\begin{document}

\maketitle
\begin{abstract}
Observation decimation is frequently employed in state estimation to reduce sensing, communication, 
and computational requirements, but decreasing the measurement assimilation frequency increases estimation uncertainty. 
Selecting an appropriate observation decimation factor therefore requires accurately predicting the resulting 
estimator performance. 
While the discrete algebraic Riccati equation (DARE) provides the steady-state estimation-error covariance 
for standard linear time-invariant Kalman filters, it is not directly applicable to estimators employing 
decimated measurement updates. Existing approaches address this limitation through lifted system representations 
or periodic Riccati equation formulations, both of which incur additional computational complexity. 
This paper presents a covariance-constrained observation decimation (CCOD) framework that reformulates the 
DARE inputs using equivalent decimated system and process-noise matrices that capture covariance 
growth between measurement updates. 
The proposed reformulation enables direct prediction of the steady-state estimation-error covariance through a 
single DARE evaluation without increasing the system dimension or solving coupled periodic Riccati equations. 
The resulting covariance prediction is used to determine the maximum observation decimation factor that satisfies 
a prescribed estimation uncertainty bound. Validation using a high dimensional linear time-invariant system 
and a space object tracking application demonstrates that the proposed approach accurately predicts steady-state 
estimator performance while reducing the measurement assimilation frequency required to satisfy specified 
covariance constraints.
\end{abstract}
\section{Introduction}

Standard estimation frameworks can be modified to execute an observation update step after a sequence of consecutive state propagation steps such as the decimated measurement update (DMU) estimator. DMU applications arise in response to different scenarios such as managing system resource limitations including Size, Weight, and Power (SWaP) constraints, intermittent observations due to environmental conditions (e.g., aerospace robotics, GPS-denied scenarios, or satellite orbiting dynamics); or communication limitations caused by latency and bandwidth constraints on remote sensing applications  \cite{nesnas2021autonomy}. 

In determining the appropriate decimation factor, it is necessary to determine the impact that dropped observations have on the quality of the state estimate relative to requirements in a particular application. Performance of a candidate estimator can be established through Monte Carlo simulation, but this is computationally expensive and not appropriate for determining the decimation at run-time. Kalman filters for linear, time-invariant systems with constant process and measurement noise covariances reach a steady-state where the estimation-error covariance converges to a constant value \cite{simon2006optimal}. The steady-state Kalman filter defines the long-term, predictable estimator performance providing a more efficient metric for assessing the enduring performance of a system compared to Monte Carlo simulations. 

While the discrete algebraic Riccati equation (DARE) yields a solution to the steady-state covariance of a standard Kalman filter, it is not directly applicable to systems subject to a DMU. To address this, Yang \cite{yang2018efficient} utilizes a lifting method. Lifting allows using the standard DARE for a periodic system by stacking the system dynamics over a period, which augments the dimensionality of the system matrices and increases computational cost. Alternatively, Sagfors and Toivonen \cite{saagfors1998h} obtain a solution for a periodic system by using the discrete-time periodic Riccati equation (DPRE). However, DPRE converts the problem into a sequence of coupled equations where the solution of one equation requires the solution of the next; consequently, the computational cost grows as the decimation factor grows.  

The main contribution of this work is a reformulation of the DARE inputs that enables a direct computation of the steady-state covariance of an estimator subject to DMU, while avoiding increased system dimensionality or coupled periodic Riccati equations. The proposed covariance-constrained observation decimation (CCOD) approach replaces the original DARE system and process noise matrices with equivalent decimated matrices that capture covariance growth between measurement updates. These equivalent system and process noise matrices allow the steady-state solution from a single evaluation of the DARE. The reformulated DARE inputs are used to predict the steady-state estimator-error covariance of an arbitrary linear system subject to a DMU Kalman filter implementation. Finally, the prediction can be used to establish the minimum measurement data assimilation frequency required to maintain an estimation-error covariance within a predefined threshold. The remainder of this paper is organized as follows: Section 2 reviews steady-state covariance theory and CCOD filter decimation; Section 3 outlines the experimental setup; Section 4 presents the results; and Section 5 concludes the work.

\section{Steady-State Covariance}\label{sec:steady_state_covariance}

A key step for implementing the DMU is to predict the steady-state covariance of a filter. For linear, discrete time-invariant (LDTI) systems with constant process and measurement covariances ($Q$ and $R$), the Kalman gain $K$ of a linear Gaussian system  approaches a steady-state value if the system satisfies specific structural properties. Specifically, Simon shows the Kalman filter's \textit{a priori} state covariance ($P^{-}_k$) for large $k$ approaches a unique, steady-state value if the system pair $(A,B)$ is stabilizable and the pair $(A,H)$ is detectable \cite{simon2006optimal}.

When these stabilizability and detectability conditions are met, the steady-state covariance is obtained as the unique stabilizing solution to the discrete algebraic Riccati equation (DARE)\cite{simon2006optimal}: 
\begin{equation*}
    P_{\infty} = A P_{\infty} A^{\intercal} - A P_{\infty} H^{\intercal} (H P_{\infty} H ^{\intercal} + R)^{-1} H P_{\infty} A^{\intercal} + Q 
    \label{eq:DARE}
\end{equation*}
where $P_{\infty}$ is the steady-state \textit{a priori} covariance, $A$ is the system matrix, $H$ is the observation model, and $Q$ is the covariance process noise. Furthermore, satisfying these stabilizability and detectability conditions guarantees the $P_{\infty}$ will yield a stable estimator. Conversely, if these structural conditions are violated, a stale steady-state solution does not exist, and the estimator error covariance will grow unbounded.

\section{Filter Decimation}
\begin{figure}[h]
    \centering
\begin{tikzpicture}[
    node distance=2cm,
    latent/.style={circle, draw, thick, minimum size=0.3cm},
    obs/.style={circle, draw, thick, minimum size=0.3cm,fill=gray!20},
    arrow/.style={->, thick},
    dotted_arrow/.style={->, thick, dotted}
    ]

% Nodes
    \node[latent] (X1) {};
    \node[below right=0.01cm of X1] (X1_label) {$\mathbf{x}_1$};
    \node[obs, below=0.4cm of X1] (o1) {};
    \node[below =0.01cm of o1] (o1_label) {$\mathbf{y}_{1}$};
    
    \node[latent, right=1.0 cm of X1] (X2) {};
    \node[below right=0.01cm of X2] (X2_label) {$\mathbf{x}_{2}$};
    \node[latent, above right =0.7cm of X1] (u1) {};
    \node[above=0.01cm of u1] (u1_label) {$\mathbf{u}_{1}$};
    % \node[obs, below=0.4cm of X2] (o2) {};
    % \node[below =0.01cm of o2] (o2_label) {$\mathbf{y}_{i+1}$};
    
    \node[latent, right=1.0cm of X2] (X3) {};
    \node[below right=0.01cm of X3] (X3_label) {$\mathbf{x}_{k \mathrm{\ mod \ } d = 0}$};
    \node[latent, above right = 0.7 cm of X2] (u2) {};
    \node[above=0.01cm of u2] (u2_label) {$\mathbf{u}_{k-1}$}; 
    \node[obs, below=0.4cm of X3] (o3) {};
    \node[below =0.01cm of o3] (o3_label) {$\mathbf{y}_{k}$};

    \node[latent, right=1.2cm of X3] (X4) {};
    \node[below right=0.01cm of X4] (X4_label) {$\mathbf{x}_{N}$};
    \node[latent, above left= 0.7 cm of X4] (u3) {};
    \node[above=0.01cm of u3] (u3_label) {$\mathbf{u}_{N-1}$};  
    % \node[obs, below=0.4cm of X4] (o4) {};
    % \node[below =0.01cm of o4] (o4_label) {$\mathbf{y}_{p}$};
        
    % Arrows
    \draw[arrow] (X1) -- (X2);
    \draw[dotted_arrow] (X2) -- (X3);
    \draw[dotted_arrow] (X3) -- (X4);

    \draw[arrow] (X1) -- (o1);
    % \draw[arrow] (X2) -- (o2);
    \draw[arrow] (X3) -- (o3);
    % \draw[arrow] (X4) -- (o4);
 
    \draw[arrow] (u1) -- (X2); 
    \draw[arrow] (u2) -- (X3); 
    \draw[arrow] (u3) -- (X4); 
\end{tikzpicture}
\caption{PGM of decimated system.}
\label{fig:decimation_pgm}
\end{figure}
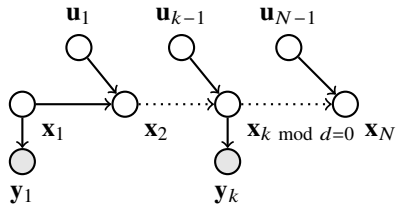

We aim to determine the steady-state $P^{-}_k$ of an estimator subject to a DMU. To illustrate our definition of a DMU, Figure \ref{fig:decimation_pgm} shows the probabilistic graphical model (PGM) of a DMU system where the observations $\mathbf{y}_k$ are assimilated only when $k$ (mod $d$) = 0, for $k = 1,2,\dots,N$. Here, $d$ represents the number of consecutive time steps that the system does not perform a measurement update with $\mathbf{u}_k$ as the control input vector and $\mathbf{x}_k$ is the state vector.

\begin{figure}[h]
    \centering
    \includegraphics[scale=0.5]{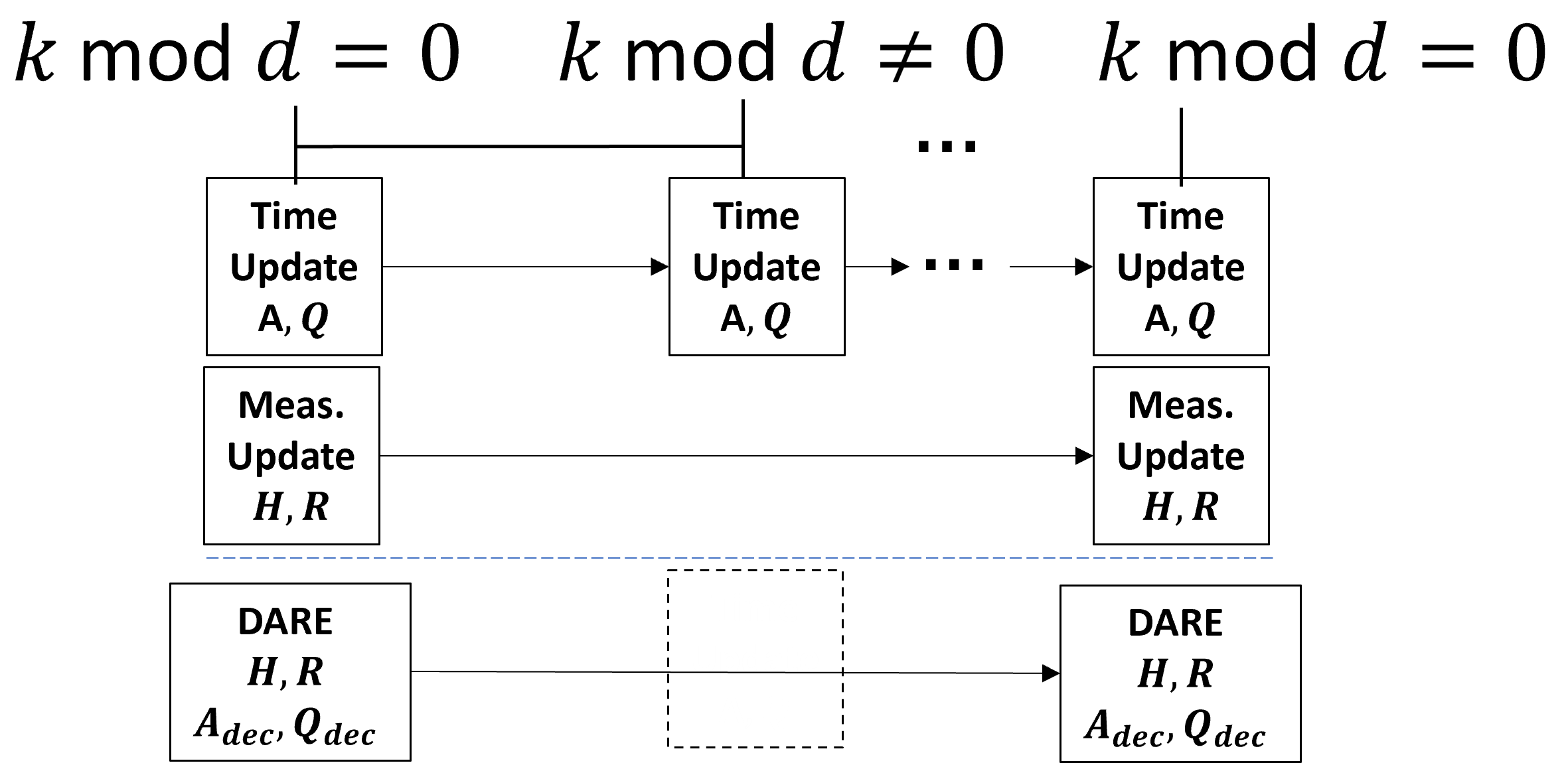}
    \caption{$A$, $Q$, $H$, and $R$ matrices for a decimated system and $A_{\mathrm{dec}}$ and $Q_{\mathrm{dec}}$ for the DARE.}
    \label{fig:decimation}
\end{figure}

The DARE provides a steady-state solution for a non-decimated discrete-time system. However, directly applying the original system matrices $A$ and $Q$ to a DMU estimator does not capture covariance growth during skipped measurement epochs. To capture the effect of decimation, we develop modified $A_{\mathrm{dec}}$ and $Q_{\mathrm{dec}}$ matrices, which enable a DARE-based solution for any decimation value $d$. As shown in Figure \ref{fig:decimation}, these modified matrices $A_{\mathrm{dec}}$ and $Q_{\mathrm{dec}}$ must capture the state evolution and covariance growth through epochs which are present in the original system but absent in the decimated system. Notably, the $H$ and $R$ (measurement noise covariance) matrices remain unmodified, as they are not involved in the time update. The decimation reformulation allows for a direct steady-state solution while maintaining the original system dimensionality. 

\textbf{Proposition 1 (Equivalent Decimated System).}

Consider a LDTI system subject to a DMU with integer decimation factor $d$. The equivalent decimated system matrices are given by: 
% To develop expressions for $A_{\mathrm{dec}}$ and $Q_{\mathrm{dec}}$, we consider a LDTI system subject to a DMU with an integer decimation factor $d$. AT every measurement update instant, the system can be represented with equivalent decimated system matrix $A_{\mathrm{dec}}$ and equivalent process noise matrix $Q_{\mathrm{dec}}$:
\begin{subequations}
    \begin{equation}
        A_{\mathrm{dec}} = A^d
    \label{eq:A_dec}
    \end{equation}
    \begin{equation}
        Q_{\mathrm{dec}} = \sum_{j=0}^{d-1} A^j Q (A^j)^{\intercal}
    \label{eq:Q_dec}
    \end{equation}
    \label{eq:decimated_equations}
\end{subequations}
such that the covariance propagation across $d$ consecutive prediction steps without intermediate measurement updates satisfies: 
% These matrices characterize the discrete-time evolution of $P^{-}_k$ over $d$ consecutive steps without intermediate measurement updates:

\begin{equation*}
    \begin{split}
    P^{-}_{k+d} &= A^d P_{k}^+ (A^d)^{\intercal} + \sum_{j=0}^{d-1} A^j Q (A^j)^{\intercal} \\
                &= A_{\mathrm{dec}} P^{+}_{k} A_{\mathrm{dec}}^{\intercal} + Q_{\mathrm{dec}}              
    \end{split}
\label{eq:prior_covariance_decimated}
\end{equation*}
substituting these $A_{\mathrm{dec}}$ and $Q_{\mathrm{dec}}$ into the standard DARE yields the decimated DARE: 

\begin{equation*}
    P_{\infty_{\mathrm{dec}}} = A_{\mathrm{dec}} P_{\infty_{\mathrm{dec}}} A_{\mathrm{dec}}^{\intercal} - A_{\mathrm{dec}} P_{\infty_{\mathrm{dec}}} H^{\intercal} (H P_{\infty_{\mathrm{dec}}} H ^{\intercal} + R)^{-1} H P_{\infty_{\mathrm{dec}}}A_{\mathrm{dec}}^{\intercal} + Q_{\mathrm{dec}} 
    \label{eq:DARE_decimated}
\end{equation*}
where $ P_{\infty_{\mathrm{dec}}}$ is the steady-state \textit{a priori} covariance under decimation.

\textit{Proof.}

The \textit{a priori} state estimate propagation of a standard discrete-time Kalman filer over a single step is governed by \cite{thrun2005probabilistic}:
 
\begin{equation*}
\hat{\mathbf{x}}_{k+1}^- = A \hat{\mathbf{x}}^+_k + B \mathbf{u}_{k} 
\label{eq:prior_state_kplus1}
\end{equation*}
where $\hat{\mathbf{x}}_{k+1}^+$ is the \textit{a posteriori} state estimate from the previous step and $B$ is the control-input matrix. Recursively applying this single-step propagation equation over an arbitrary interval of $i$ subsequent steps yield the standard multi-step state trajectory expansion documented in the literature \cite{simon2006optimal}: 

\begin{equation}
    \hat{\mathbf{x}}_{k+i}^- = A^i \hat{\mathbf{x}}_{k}^+ + \sum_{j=0}^{i-1} A^j B \mathbf{u}_{k+i-1-j}
\label{eq:arbitrary_step_state}
\end{equation}
Equation \ref{eq:arbitrary_step_state} represents the deterministic, un-updated state estimate trajectory across the decimation gap.  
While the state estimate propagates deterministically via the multi-step equation, the uncertainty of this estimate grows due to stochastic disturbances. The effect of the process noise is captured by tracking the error covariance matrix. Assuming a constant process noise $Q$, the standard one-step propagation of the state covariance is given by: 
\begin{equation*}
    P^{-}_{k+1} = A P^{+}_{k} A^{\intercal} + Q 
    \label{eq:prior_covariance_kplus1}
\end{equation*}
evaluating the covariance over two time steps without an intermediate update means setting $P^+_{k+1} = P^-_{k+1}$. Substituting the one-step propagation into the next step yields:   

\begin{equation*}
    \begin{split}
    P_{k+2}^{-} & = A (P^-_{k+1}) A^{\intercal} + Q \\
                & = A (A P_k^+ A^{\intercal} + Q) A^{\intercal} + Q \\
                & = AA P_k^+  A^{\intercal} A^{\intercal} + A Q  A^{\intercal} + Q
    \end{split}
\label{eq:prior_covariance_kplus2}
\end{equation*}
generalizing to an arbitrary number of prediction steps $i$, the \textit{a priori} covariance at time $k+i$ is:    

\begin{equation*}
    P_{k+i}^{-} = A^i P_{k}^+ (A^i)^{\intercal} + \sum_{j=0}^{i-1} A^j Q (A^j)^{\intercal}
\label{eq:arbitrary_step_prior_covariance}
\end{equation*}
setting $i = d$ establishes the representations for $A_{\mathrm{dec}}$ and $Q_{\mathrm{dec}}$ shown previously in Equation \ref{eq:decimated_equations}.

\section{Experiment and Case Study}

We implement an experiment and a case study that leverage the steady-state covariance prediction from the DARE to determine the maximum permissible decimation factor for a Kalman filter without exceeding a user-specified covariance threshold. In these particular experiments, the implemented constraint threshold is governed by the maximum variance among all individual states, ensuring that no diagonal element of the steady-state covariance matrix exceeds a maximum allowable threshold. However, this framework is generalizable; any scalar functional of the covariance matrix can be substituted as the limiting criterion. 

\subsection{Arbitrary System Experiment}
For this experiment, a scalable, synthetic linear system of arbitrary state dimension $\mathbf{x} \in \mathbb R^{n}$ is constructed. Utilizing a generalized synthetic system enables precise control over structural properties, specifically controllability, observability, and stability which are difficult to isolate in a fixed physical system model. To thoroughly test the boundaries of the decimated filter, the system is designed to exhibit complex dynamics, including highly oscillatory and marginally stable modes. Analyzing the complex dynamics requires precise placement of the system eigenvalues. The complete procedure for generating an arbitrary, structurally controlled system matrix is detailed in the Appendix. 

We generate a system matrix $A$ of size $n=20$ using Equation \ref{eq:arbitrary_system} where all eigenvalues of $A$ are complex. The values of $\sigma_k$ and $\omega_k$ are selected randomly between 0.6 and 0.7. Limiting $\sigma_k$ and $\omega_k$ to the interval $[0.6,0.7]$ will result in the magnitude of the resulting complex eigenvalues to always be less than 1 $(0.84 < \rho < 0.99 )$. This ensures the dynamics are always relatively close to the system stability limit (for an asymptotically stable discrete time invariant linear system, the magnitude of its eigenvalues must be less than 1 \cite{simon2006optimal}). 

The input matrix $B$ is defined such that only half of the states receive direct input. We arbitrarily select which states receive a direct input, where those states receiving input are assigned unity gain. From the partially controllable system, the controllability matrix is constructed per \cite{simon2006optimal}:

\begin{equation*}
\mathcal{P}=\begin{bmatrix}
    B  &  A B & A^2 B & \dots &  A^{n-1} B
\end{bmatrix}
\end{equation*}
here, $n$ represents the dimension of the state vector, and a full rank controllability matrix $\mathcal{P}$ confirms the pair $(A,B)$ is fully controllable. 

The output matrix $H$ is defined such that only half of the states are directly observed. The observed states are arbitrarily selected. The system observability matrix $\mathcal{O}$ is constructed per \cite{simon2006optimal}:
\begin{equation*}
\mathcal{O}=\begin{bmatrix}
    H^\intercal  & (H A)^\intercal & (H A^2)^\intercal & \dots & (H A^{n-1})^\intercal 
\end{bmatrix}^\intercal
\end{equation*}
a full rank observability matrix $\mathcal{O}$ confirms that the pair $(A,H)$ is fully observable. Because full controllability and observability strictly imply that the system is also stabilizable and detectable, these structural properties satisfy the convergence conditions outlined by Simon \cite{simon2006optimal}. The structural properties ensure the existence of a unique, positive-semidefinite steady-state covariance solution $P_{\infty}$.

The measurement noise is set equal for all the measurement states to be Gaussian with $\nu \sim \mathcal{N}(0,0.1)$. For the states under direct input, we consider a zero input with Gaussian noise given by $\omega \sim \mathcal{N}(0,1)$ which yields $Q = B \Sigma_{\omega}B^\intercal$.

\subsection{Space Object Tracking Case Study}

To evaluate the proposed CCOD approach using an independent, real-world data verification benchmark, a relative space object tracking problem is implemented. This case study models a proximity operations tracking scenario where a chaser vehicle maintains a local state estimate of a target. To isolate the estimator’s performance and ensure compliance with the Linear Time-Invariant (LTI) constraints required by the CCOD framework, the tracking filter cannot embed non-linear orbital dynamics, nor can it process irregular, time-varying sampling intervals. To bridge these non-linear dynamics and time-varying sampling intervals constraints, a constant-interval simulation pipeline was constructed utilizing raw historical Two-Line Element (TLE) records for the International Space Station (ISS) from January 1, 2026 \cite{SpaceTrackGP}. 

This simulation is achieved using the Skyfield astrodynamics library \cite{2019ascl.soft07024R},\cite{Rhodes2019Skyfield}. To handle irregular gaps, the pipeline dynamically scans the database at each constant time step $\Delta t$, loading the most recent chronological TLE into a Simplified General Perturbations (SGP4) engine \cite{Hoots1980}. SGP4 uses these discrete historical parameters as updated initial conditions to analytically simulate and sample the absolute inertial state forward \cite{Vallado2006}.

To evaluate the LTI tracking performance, a secondary target object is initialized with a stable in-track phase separation by slightly shifting its Mean Anomaly relative to the ISS, and both absolute tracks are propagated forward independently using the SGP4 engine over a multi-day window. Because the global Earth-Centered Inertial (ECI) frame is highly non-linear, the absolute states are continuously transformed into a local relative Radial, Along-Track, Cross-Track (RTN) frame whose origin is attached to the ISS along a circular nominal reference orbit ($r_{\mathrm{nominal}} = 6771.0 \ \mathrm{km}$). By isolating small relative displacements around this localized, rotating origin, the dominant non-linear central gravity terms cancel out, yielding the linearized Clohessy-Wiltshire equations of motion. This transformation utilizes a standard frame rotation for positions and incorporates the required kinematic Coriolis transport corrections ($\mathbf{\omega} \times \mathbf{r}$) for relative velocities to maintain tracking fidelity.

This state mapping yields the continuous LTI system governed by the linearized Clohessy-Wiltshire (CW) equations \cite{curtis2019orbital}. Since this work implements a first-order discrete Euler integration scheme, the constant system matrix $A \in \mathbb{R}^{6 \times 6}$ is explicitly defined by: 

\begin{equation*}
    A = \begin{bmatrix}
        1 & 0 & 0 & \Delta t & 0 & 0 \\
        0 & 1 & 0 & 0 & \Delta t & 0  \\
        0 & 0 & 1 & 0 & 0 & \Delta t  \\
        3 n^2 \Delta t & 0 & 0 & 1 & 2n \Delta t & 0 \\
        0 & 0 & 0 & -2n \Delta t & 1 & 0 \\
        0 & 0 & -n^2 \Delta t & 0 & 0 & 1
    \end{bmatrix}
\end{equation*}
where $n = (\frac{\mu}{r^3_{\mathrm{nominal}}})^{\frac{1}{2}}$ represents the orbital mean motion. 

% \subsubsection{Estimator Parameters and Initial Conditions}

The discrete-time estimation state vector is defined by the relative kinematics $\mathbf{x} = \begin{bmatrix}
        x & y & z & \dot{x} & \dot{y} & \dot{z}
    \end{bmatrix}^\intercal$. The observation model tracks the position vector $\mathbf{y} = \begin{bmatrix}
        x & y & z 
    \end{bmatrix}^\intercal$.
% The measurement noise covariance matrix $R$ is structured to bound the order-of-magnitude structural uncertainty inherent to public TLE catalog data, which historically demonstrates kilometer-level tracking and prediction accuracy due to truncated force models and sparse observation geometries \cite{vallado2007analysis}:

The measurement noise covariance matrix $R$ is set to 10 m$^2$, establishing a conservative variance baseline for vision-based position measurements.  

\begin{equation*}
    R = \mathbf{I}_{3 \times 3} 0.01 \ \mathrm{km}^2.
\end{equation*}

The process noise covariance matrix \(Q\) is parameterized to capture the unmodeled nonlinearities, orbital perturbations, and Earth oblateness effects (\(J_{2}\)) omitted by the linear Clohessy-Wiltshire state transition model relative to the SGP4 truth engine:

\begin{equation*}
    Q = \begin{bmatrix}
         \mathbf{I}_{3 \times 3} 10^{-12} \ \mathrm{km}^2 & \mathbf{0}_{3 \times 3} \\
         \mathbf{0}_{3 \times 3} & \mathbf{I}_{3 \times 3} 10^{-14} \frac{\ \mathrm{km}^2}{s^2}
    \end{bmatrix} \Delta t
\end{equation*}

\subsection{Finding the Maximum Measurement Decimation Factor}
To determine the maximum permissible decimation factor, $P_{\infty_{\mathrm{dec}}}$ is iteratively evaluated via the decimated DARE for increasing values of $d$. This iterative search terminates when any diagonal element of $P_{\infty_{\mathrm{dec}}}$ exceeds the user-specified threshold $P^-_{\mathrm{desired}}$ and the decimation factor that satisfies the condition is obtained. Finally, a discrete-time Kalman filter subject to a DMU is then implemented with the obtained decimation factor to evaluate the evolution of $P^-_k$. 
\section{Results}

We now demonstrate the performance of the proposed CCOD implementation using a controlled experiment and a case study: a generalized arbitrary linear system and a relative space object tracking case. The covariance behavior is tracked to verify the performance of the CCOD approach as the estimator's covariance subject to a decimation factor $d$ converges to a steady state value. 

\subsection{Arbitrary System}
\begin{figure}[h]
    \centering
    \includegraphics[width=\textwidth]{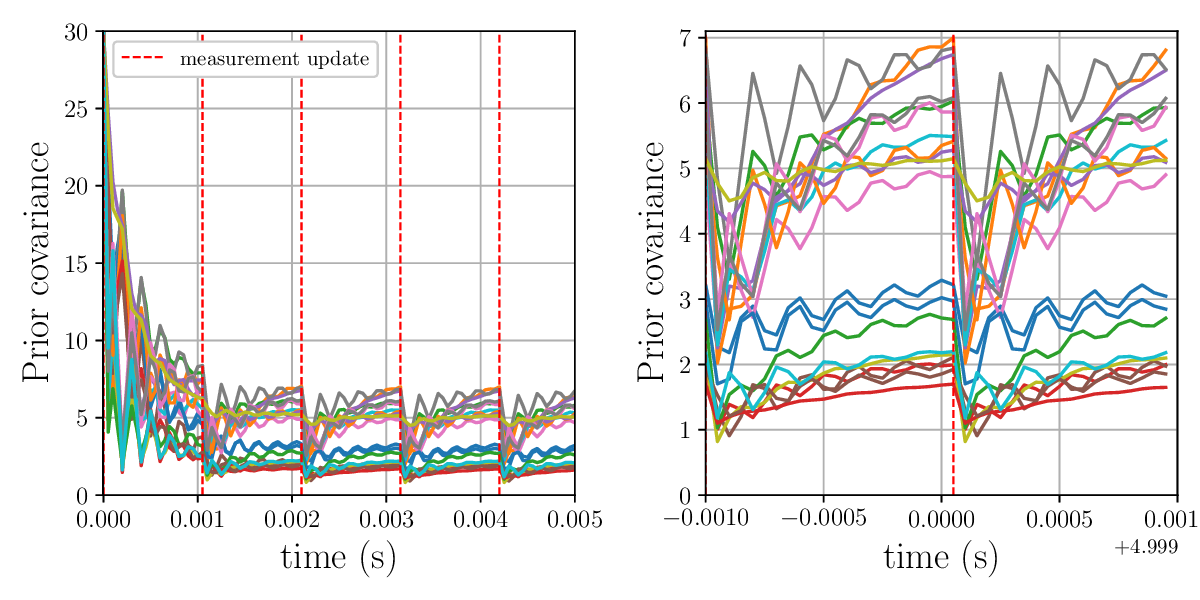}
    \caption{Time history of $P^-_k$ variance terms associated with all the states for a Kalman filter implementation subject to a DMU with decimation factor of 21. Simulation time is 5 seconds, time-step used is 5 $\times$ 10$^{-5}$ seconds. Left: Initial 0.005 seconds of simulation. Right: Final 0.002 seconds of simulation}
    \label{fig:prior_covariance_experiment}
\end{figure}

Figure \ref{fig:prior_covariance_experiment} illustrates the time history of $P^-_k$ for an arbitrary 20-state linear system evaluated via a CCOD discrete-time Kalman filter implementation. A maximum allowable variance threshold of 7.0 (dimensionless, as the system is arbitrary) is specified, yielding a permissible decimation factor of $d=21$. The vertical dotted red lines in the plot indicate the discrete time-steps when a measurement update occurs. 

Within the initial 0.025 seconds of the simulation, the state variances immediately bifurcate into two distinct groups. One group settles above 17 and the other drops below 10. This bifurcation occurs because the directly measured states are instantly reduced during the initial measurement update at $t=0$ seconds, while all covariance elements subsequently propagate through the discrete-time dynamics via single-step time updates.

Due to the dynamics coupling, the two variance groups intermix during the subsequent propagation steps. After a 5-second simulation duration, the maximum steady-state variance element converges to 6.9995, validating the modified DARE prediction by remaining just below the specified threshold of 7.0.

\subsection{Space Object Tracking}
\begin{figure}[h]
    \centering
    \includegraphics[width=\textwidth]{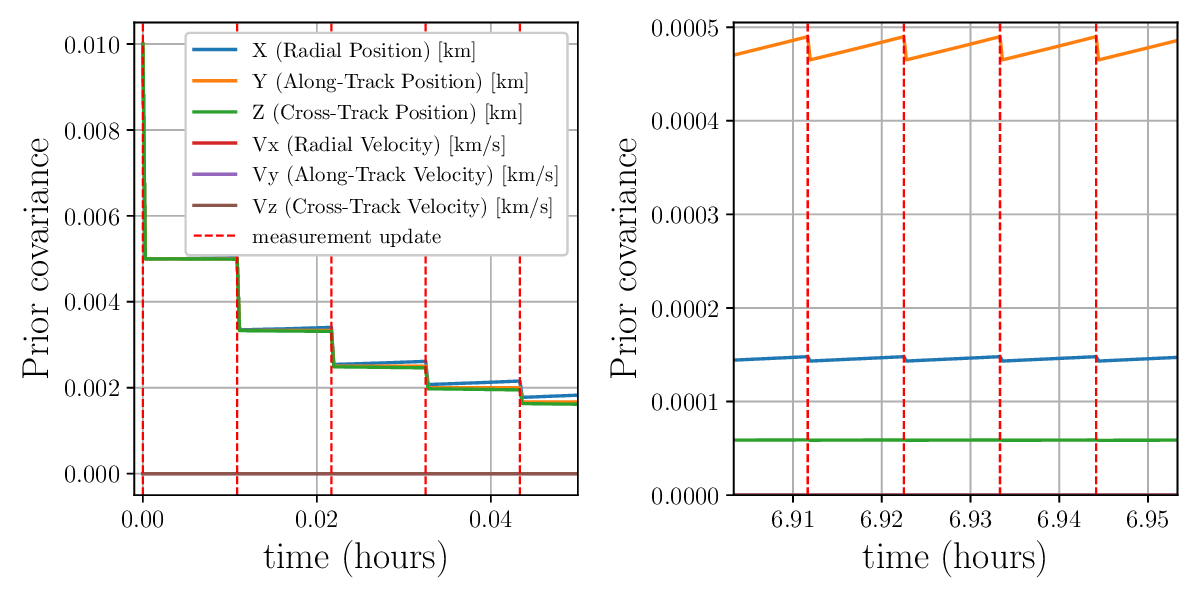}
    \caption{Time history of $P^-_k$ variance terms associated with all the states for the tracking implementation subject to a DMU with decimation factor of 39. Run time is 6.954 hours, time-step used is 1 second. Left: Initial 0.05 hours of run. Right: Final 0.05 hours of run}
    \label{fig:prior_covariance_ISS}
\end{figure}

The non-decimated system utilizes a sensor measurement frequency of 1 Hz, establishing a standard sampling rate for vision-based position measurements. Figure \ref{fig:prior_covariance_ISS} shows the time history of $P^-_k$ for the tracking test case. A maximum allowable variance threshold of 0.0005 (in this case units are either km$^2$ for relative position states or $\frac{\ \mathrm{km}^2}{\mathrm{s}^2}$ for relative velocity states) is specified. The resulting maximum decimation factor allowed is $d=39$. The maximum $P^-_k$ at the end of the run is 0.00049 km$^2$ corresponding to the along-track position confirming that the CCOD implementation is able to ensure the $P^-_k$ threshold is not exceeded after implementing a decimation factor of $d$.

\section{Conclusion}

A DMU reduces the amount of data an estimator needs to assimilate at the expense of increased estimator covariance. The DARE predicts the steady-state covariance of non-decimated LDTI system estimator implementations where the estimator covariance converges. Obtaining a steady-state covariance prediction for a DMU estimator implementation is more challenging and requires reformulation of the DARE inputs that account for the covariance growth that happens in intervals between measurement updates due to process uncertainty. 

To account for a covariance growth without observations, we developed expressions for the system matrix $A_{dec}$ and process noise matrix $Q_{dec}$ subject to decimation. These expressions address a critical covariance prediction implementation step and are summarized in Equation \ref{eq:decimated_equations}. We also provided proof that these new expressions for the decimated system and process noise matrices are able to predict the steady-state covariance of a DMU estimator implementation when used as inputs to the DARE. 

To test our CCOD approach, we implement an arbitrary large LDTI system and a space object tracking experiments. In both cases, a DMU estimator implementation successful maintains the steady state covariance below a desired threshold while maximizing the level of measurement update decimation factor, reducing the amount information the estimator needs to assimilate in the form of sensor data.    
Results are shown for linear systems and potential extensions to nonlinear systems are left for future work.

\clearpage
\appendix
\renewcommand{\theequation}{A\arabic{equation}}
\setcounter{equation}{0}
\section*{Appendix: Structurally controlled System Matrix Design} \label{sec:system_matrix_design}

To construct a scalable, real-valued system matrix $A \in \mathbb{R}^{n\times n}$ with precise control over stability profiles, damping properties, and the distribution of oscillatory vs. non-oscillatory modes, an eigen-decomposition framework is utilized \cite{petersen2008matrix,kreyszig2025advanced}: 

\begin{equation}
    A = V \Lambda V^{-1}
    \label{eq:arbitrary_system}
\end{equation}
where $A$ is a square matrix, $V$ is a square, orthogonal eigenvector matrix, and $\Lambda \in \mathbb{R}^{n \times n}$ is the composite, block-diagonal modal matrix containing the desired discrete-time properties. Utilizing an orthogonal eigenvector matrix guarantees an ideal matrix condition number of 1, eliminating floating-point inversion errors and allowing replacement of $V^{-1}$ with the matrix transpose $V^\intercal$. 

\subsection*{Construct the Orthogonal Eigenvector Matrix}
A random matrix $V_0 \in \mathbb{R}^{n \times n}$ is initialized with elements drawn randomly. A QR decomposition is performed on this matrix to enhance numerical robustness, ensuring a condition number of 1.

\subsection*{Assemble the Modal Matrix $\Lambda$ for Mixed Dynamics}
Let $N_r$ be the number of desired purely real eigenvalues and $N_c$ be the number of desired complex conjugate pairs, such that the total state dimension satisfies $n = N_r + 2N_c$. The global modal matrix $\Lambda$ is formed as a block-diagonal composition of scalar real eigenvalues and $2 \times 2$ real-valued blocks representing complex modes: 

\begin{equation*}
 \Lambda = \text{blockdiag}(\lambda_1, \dots, \lambda_{N_r}, \Lambda_1, \dots, \Lambda_{N_c})
\end{equation*}
The stability limits and modal frequencies are governed by specifying the geometric boundaries of these entries relative to the discrete-time unit circle. 

\subsubsection*{Real Blocks (Lowercase $\lambda_i$)}
For non-oscillatory modes, the purely real eigenvalues are represented by scalars $\lambda_i \in \mathbb{R}$. These are assigned directly to the diagonal of $\Lambda$. To guarantee asymptotic stability, these elements are bounded within the unit circle ($|\lambda_i| < 1$). Marginally stable modes are placed exactly on the boundary ($|\lambda_i| = 1$), and unstable modes are placed outside ($|\lambda_i| > 1$).

\subsubsection*{Complex Blocks (Uppercase $\Lambda_k$)}
For oscillatory modes, each complex conjugate pair is defined by a complex scalar $\gamma_k = \sigma_k + j\omega_k \in \mathbb{C}$ (where $j = (-1)^{\frac{1}{2}}$). Each pair is mapped to a real-valued $2 \times 2$ sub-block $\Lambda_k$ via the rotation-scaling theorem \cite{margalit2017interactive}: 
\begin{equation}
    \Lambda_k = \begin{bmatrix}
        \mathrm{Re}(\gamma_k) & \mathrm{Im}(\gamma_k) \\
        -\mathrm{Im}(\gamma_k) & \mathrm{Re}(\gamma_k)  \\
    \end{bmatrix} = \begin{bmatrix}
        \sigma_k & \omega_k \\
        -\omega_k & \sigma_k  \\
    \end{bmatrix}
    \label{eq:jordan_blocks}
\end{equation}
where the discrete-time pole magnitude $\rho = (\sigma_k^2 + \omega_k^2)^{\frac{1}{2}}$ dictates the system stability. For damped oscillations, $\rho < 1$; for sustained marginal oscillations, $\rho = 1$; and for unstable dynamics, $\rho > 1$. 

\subsection*{Compute the Real System Matrix}
The final system matrix is evaluated directly via the matrix triple product in Equation \ref{eq:arbitrary_system}. Because $\Lambda$ utilizes real-valued coordinate transformations ($\Lambda_k \in \mathbb{R}^{2 \times 2}$), the resulting system matrix $A$ is guaranteed to be fully real-valued.

\section*{Funding Sources}
Portions of this work were funded by the Air Force Office of Scientific Research under award FA9550-24-1-0176.

%%%%%%%%%%%%%%%%%%%
% \clearpage
\bibliography{andres.bib}
%%%%%%%%%%%%%%%%%%

\end{document}